\documentclass[11pt, a4paper, logo, twocolumn, internal, nonumbering]{style}

\usepackage{multirow}
\usepackage{listings}
\usepackage{caption}
\usepackage{cuted}
\usepackage{subcaption}
\usepackage{amsmath}
\usepackage{booktabs}
\usepackage{xcolor}
\usepackage[style=nature]{biblatex}
\newcommand{\beginsupplement}{%
        \setcounter{tocdepth}{3}
        \setcounter{secnumdepth}{3}
        \renewcommand{\thesection}{S\arabic{section}}%
        \setcounter{table}{0}
        \renewcommand{\thetable}{S\arabic{table}}%
        \setcounter{figure}{0}
        \renewcommand{\thefigure}{S\arabic{figure}}%
     }

\title{Towards low-cost lead screening with transmission XRF}
\correspondingauthor{K.S. (kiansd@kth.se)}

\author[*,1]{Christoph Gaßner}
\author[*,2]{Juliane Reisewitz}

\author[3]{Jenna E. Forsyth}
\author[4]{Kian Shaker}

\affil[*]{Equal contributions / alphabetical order}

\affil[1]{Department of Physics, Technical University of Munich, Munich, Germany}
\affil[2]{Department of Physics, Karlsruhe Institute of Technology, Karlsruhe, Germany}
\affil[3]{Woods Institute for the Environment, Stanford University, Stanford, United States}
\affil[4]{Department of Applied Physics, KTH Royal Institute of Technology, Stockholm, Sweden}

\begin{abstract}

Human exposure to lead (Pb) is a global health concern, yet existing technologies for detecting lead in our environment remain prohibitively expensive for widespread deployment. Here we present a new concept towards lead screening using X-ray fluorescence (XRF) in an unconventional geometry we coin \textit{transmission XRF} in which the sample is placed between the source and detector. For cost reduction, we then show that $^{241}$Am found in ionizing smoke detectors is spectrally suitable for Pb L-shell XRF generation and can thus replace X-ray tubes used in conventional XRF devices. Exploring soil screening as the first application, we demonstrate with Monte Carlo simulations that a configuration with 7$\times$ $^{241}$Am sources and a standard silicon drift detector can enable screening-relevant detection limits (100 ppm Pb) in soil within practical measurement times (<30 min). We believe this concept opens a route toward low-cost and scalable XRF instrumentation for democratizing lead screening across a wide range of samples.

\end{abstract}

\begin{document}

\maketitle

\section{Introduction}

Lead (Pb) poisoning is a silent epidemic with a staggering global disease burden; no safe level of exposure is known \cite{lanphear2024lead,larsen2023global}. An estimated one third of children globally have blood lead levels above the 5 µg/dL threshold of concern, with well-established, life-long health consequences \cite{UNICEF_PureEarth_ToxicTruth_2020}. While lead exists naturally deep within the earth, safely stored from life on the surface, human mining and industrial activities release it into the biosphere \cite{luby2024removing}. Human exposure pathways include consumer products (e.g., paint \cite{o2018lead}, spices \cite{forsyth2024evidence}, cosmetics \cite{hore2024traditional}, cookware \cite{binkhorst2025potential}) as well as contaminated dust, air, water, and soil \cite{lanphear1997pathways,markus2001review, mielke1998soil}. An estimated 100 million people face lead exposure through contaminated soil at industrial sites alone, with informal used lead-acid battery (ULAB) recycling sites in low- and middle-income countries (LMICs) being major contributors \cite{hanrahan2016protecting,ericson2016global}. Even in residential areas far from industrial sites, recent estimates state that one in four U.S. households reside in areas where lead levels in soil exceed the 200 ppm threshold guidelines \cite{filippelli2024one,epa2024lead}. Protecting people requires identifying and eliminating environmental lead sources, yet large-scale screening is challenging with existing detection technologies.

Two technologies are commonly used today to detect lead in various samples: laboratory-based spectrometry (e.g., ICP-MS or ICP-OES) and X-ray fluorescence (XRF). Laboratory spectrometry offers exceptional sensitivity ($\sim$1 ppb for ICP-MS \cite{deibler2013continuing}) but relies on off-site analysis and is expensive to scale (>\$30 per sample \cite{beardsley2021method}). XRF devices enable rapid quantification of lead in various samples with great sensitivity ($\sim$10 ppm \cite{hu2017application}) and portable variants exist that are suitable for field usage \cite{sargsyan2024rapid}. However, portable XRF devices are expensive (typically >\$10,000), and while some governments and institutions can mobilize funds for such equipment, broader deployment would be enabled by lower cost instrumentation.

Motivated by this challenge, we present a concept towards low-cost and scalable XRF instrumentation by exploiting the most accessible source of X-rays available today: $^{241}$Am from ionizing smoke detectors. Surprisingly, we find that its emission spectrum is suitable for Pb detection. To leverage this weak source of X-rays, we introduce an unconventional measurement geometry that we coin \textit{transmission XRF}. As the first application, we focus on soil screening given the established screening guidelines and global concern \cite{epa2024lead}. Through simulations, we show that the transmission XRF concept with $^{241}$Am enables detection of lead in soil at threshold levels ($\sim$100 ppm) within practical measurement times (tens of minutes). We release this concept as a part of OpenXRF (\href{http://www.openxrf.org}{www.openxrf.org}) -- our initiative towards open-source and democratized XRF instrumentation for detecting heavy-element contamination in our environment.


\section{Transmission XRF with $^{241}$Am}

\begin{figure*}[ht!]
        \includegraphics[width=\textwidth]{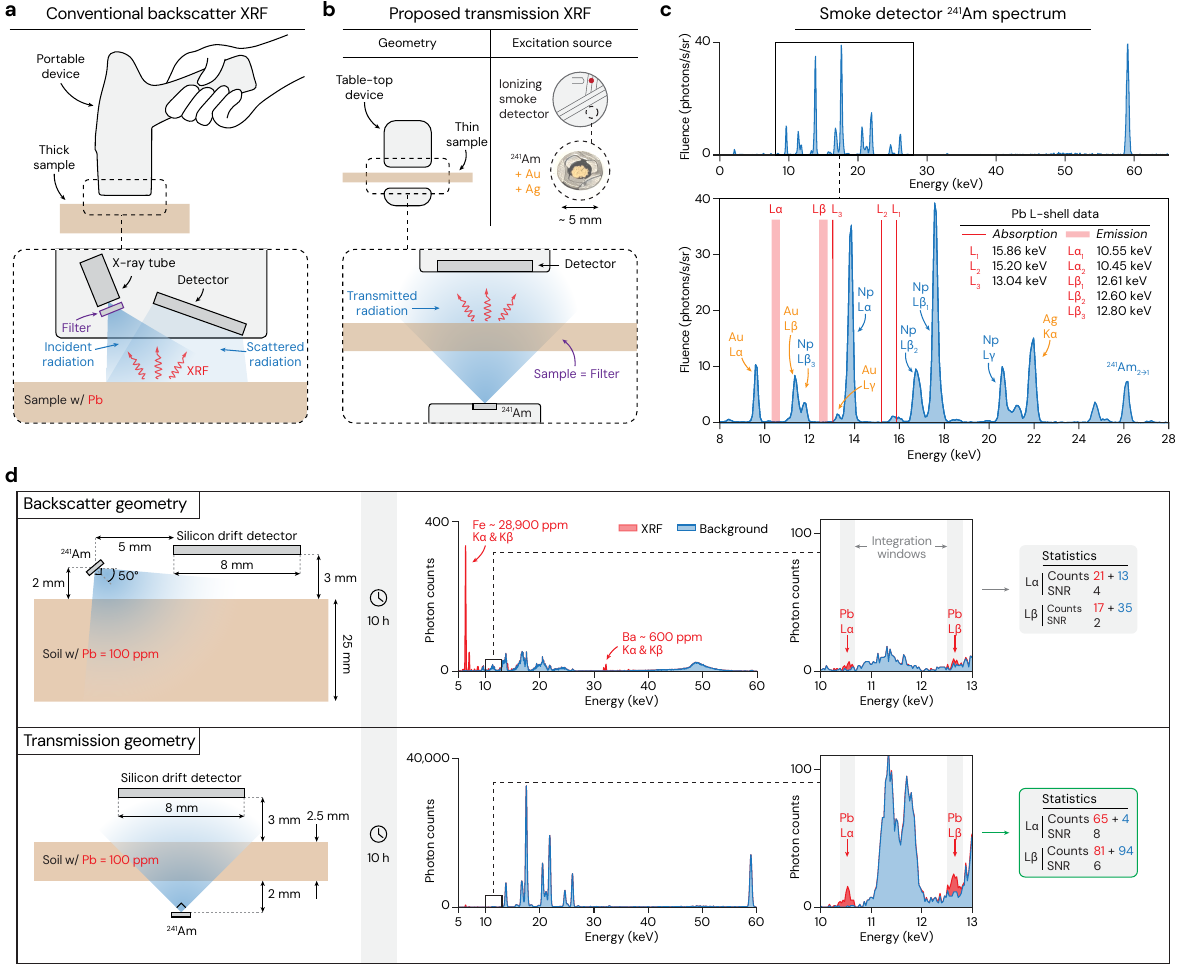}
        \caption{\textbf{Transmission XRF concept.} \textbf{a}, Conventional backscatter geometry in a portable device including an X-ray tube, filter, and detector. \textbf{b}, Our proposed transmission geometry with the sample sandwiched between the $^{241}$Am source and the detector. \textbf{c}, $^{241}$Am emission spectrum showing Np, Au, and Ag peaks (blue) as well as Pb L-shell data overlaid (red). Note that the $^{241}$Am sample in smoke detectors contains both Au and Ag from the manufacturing process \cite{belanger1979environmental}, explaining the line emissions. Np L-shell lines arise from XRF excitation of $^{237}$Np (decay product of $^{241}$Am) by the $^{241}$Am gamma emissions (mainly 59.5 and 26.3 keV).  \textbf{d}, Simulated 10-hour $^{241}$Am exposure on a soil sample, comparing backscatter and transmission geometries for 100 ppm Pb in the soil, demonstrating superior Pb detection statistics for the transmission geometry.}
        \label{fig:1}
\end{figure*}

We begin by examining conventional handheld XRF devices operating in a backscatter geometry (Fig.~\ref{fig:1}a). Here, an X-ray tube (typically a few watts, 40 kV, with a Rh/Ag/W anode) is used for excitation, and the emitted X-ray fluorescence is measured with a spectroscopic detector (typically a silicon drift detector, SDD). The emission spectrum is then filtered (typically with a thin layer of Al/Cu) to ensure that scattered radiation reaching the detector has minimal spectral overlap with the XRF of the element of interest (in our case, Pb) \cite{potts2022situ}.

\begin{figure*}[ht!]
        \includegraphics[width=\textwidth]{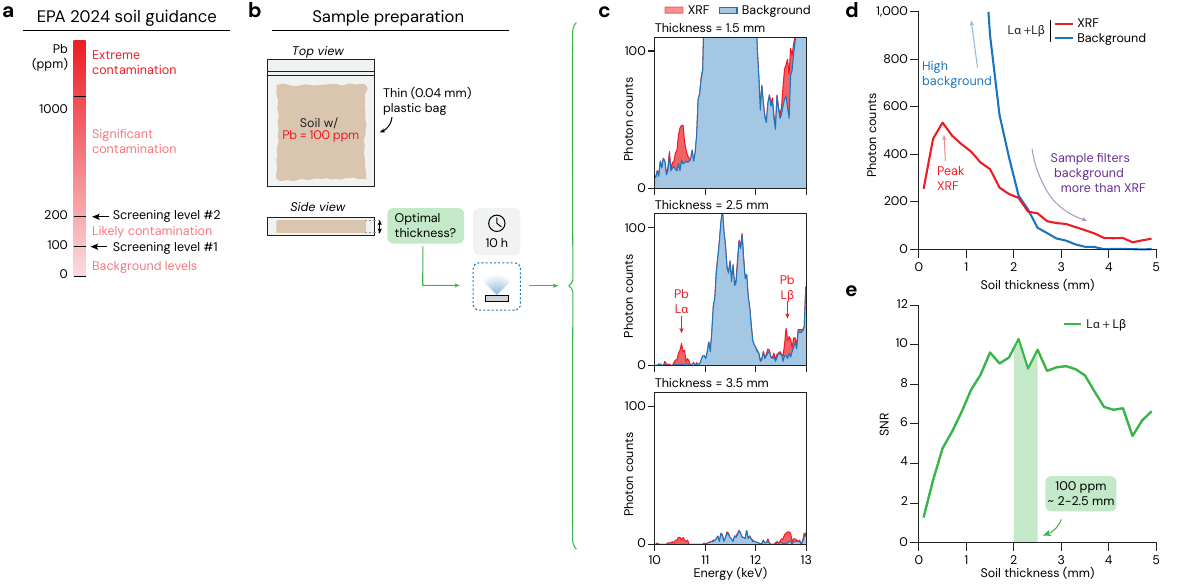}
        \caption{\textbf{Finding the optimal soil thickness.} \textbf{a}, United States Environmental Protection Agency (EPA) guidelines for soil lead levels, updated in 2024. \textbf{b}, Soil sample preparation, where we identify  the optimal thickness at 100 ppm Pb through a 10-hour $^{241}$Am simulated exposure (cf. Fig.~\ref{fig:1}d). \textbf{c}, Simulated 10-hour spectra for 1.5, 2.5, and 3.5 mm soil thickness showing Pb L$\alpha$ and L$\beta$ peaks (red) and background (blue). \textbf{d}, Integrated XRF signal (red) and background counts (blue) as a function of soil thickness, showing that the sample acts as a background filter. \textbf{e}, Signal-to-noise ratio (SNR) as a function of soil thickness for 100 ppm Pb. Results for lower (10 ppm) and higher (1000 ppm) concentrations can be found in Fig.~\ref{suppl:fig_optimal_thickness}.}
        \label{fig:2}
\end{figure*}

As we wanted to design a new concept toward low-cost XRF measurements (Fig.~\ref{fig:1}b), we decided to investigate $^{241}$Am found in ionizing smoke detectors as a potential alternative to conventional X-ray tubes, given their wide accessibility and low cost. Historically, radioisotopes have been used for XRF spectroscopy \cite{IAEA:RXFS:1970}, but the approach has been phased out in favor of X-ray tubes as the latter can be electronically controlled and avoid the concerns of handling radioactive samples. We measured the X-ray spectrum of $^{241}$Am obtained from smoke detectors (see Methods) and found, surprisingly, that the emission spectrum is highly suitable for exciting the L-shell of Pb, with the resulting L-shell XRF emission lines appearing in the gaps of the $^{241}$Am spectrum (Fig.~\ref{fig:1}c).

The primary challenge of using $^{241}$Am from smoke detectors is the extremely low fluence (Fig.~\ref{fig:1}c shows $\sim$10$^3$ photons/s/sr for a standard $\sim$37~kBq source \cite{IAEA2024SmokeDetectors}, which is over 10$^8\times$ weaker than the X-ray tubes used in conventional XRF devices). While this low activity ensures safe handling (see Suppl. Sect. ~\ref{supp:dose_calc}) along with the $^{241}$Am half-life of over 400 years for long-term stability, it negatively impacts required measurement times. To maximize the utility of $^{241}$Am in this application, we redesigned the conventional measurement setup into our proposed \textit{transmission XRF} geometry (Fig.~\ref{fig:1}b). Interestingly, in this unconventional configuration, we realized that the sample itself acts as a background filter (cf. Fig.~\ref{fig:2}a). This is particularly advantageous for our weak $^{241}$Am source, with which we cannot afford to use traditional filters, and is relevant for determining the optimal sample thickness as discussed in the next section.

How does the transmission XRF geometry compare quantitatively to the conventional backscatter approach? To explore this, we simulated a comparison (Fig.~\ref{fig:1}d, see Methods for Monte Carlo simulation details). Since the chosen distances strongly influence the results, we selected the shortest distances we deemed practically realizable for both geometries. We then simulated 10 hours of $^{241}$Am exposure and analyzed the photon spectrum detected by a hypothetical SDD placed near the sample surface. The results show that although the transmission geometry generates much higher background counts overall, the SNR at the Pb L-shell XRF peaks is substantially better. This is mainly attributed to the higher signal: the transmission geometry allows optimal placement of the $^{241}$Am source relative to both the sample and detector, enabling a much higher fraction of the Pb XRF signal to reach the detector.


\begin{figure*}[ht!]
        \includegraphics[width=\textwidth]{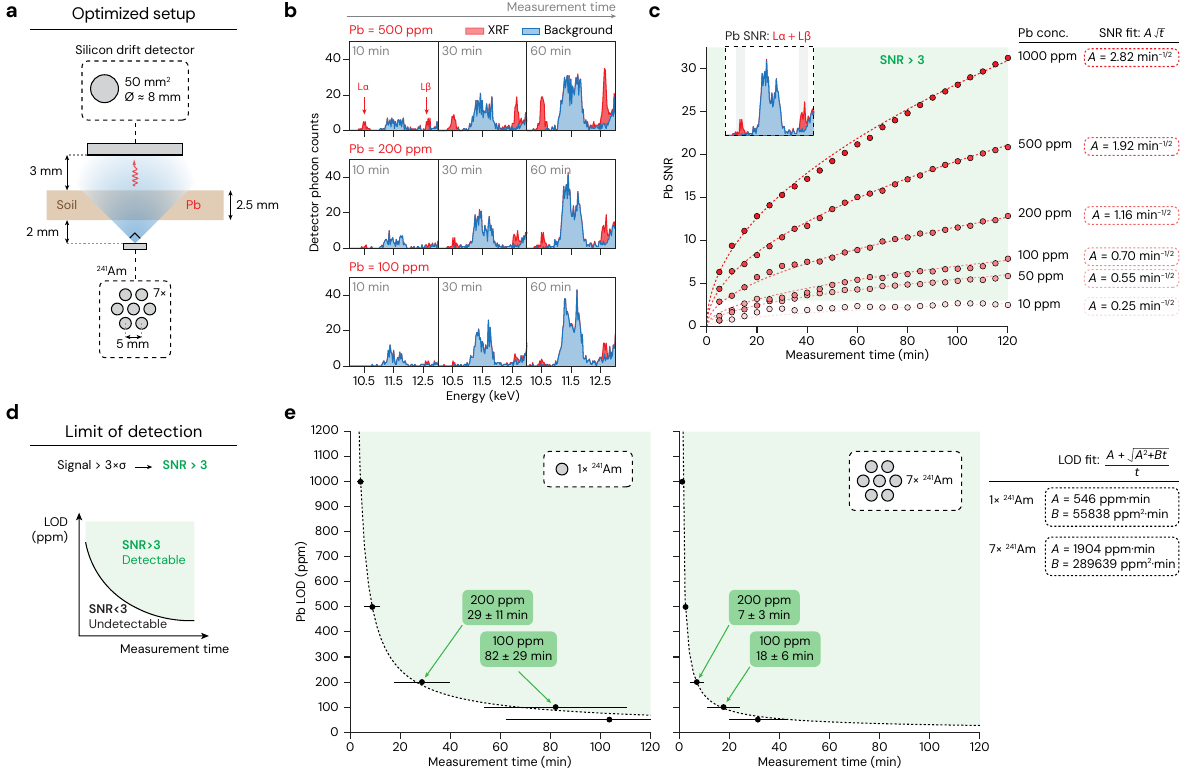}
        \caption{\textbf{Measurement time for soil screening.} \textbf{a}, Optimized arrangement with 7$\times$ $^{241}$Am sources and a 50 mm$^2$ silicon drift detector. \textbf{b}, Simulated detector spectra at example timepoints and Pb concentrations. \textbf{c}, Signal-to-noise ratio versus measurement time for the L-shell Pb XRF with the expected $\sqrt{t}$ dependence. \textbf{d}, Detection criterion schematic. \textbf{e}, Limit of detection versus measurement time for 1$\times$ and 7$\times$ $^{241}$Am configurations, with uncertainties in the measurement time calculated according to Suppl. Sect.~\ref{suppl:sec_sigma_t}. For derivations of the fit equations in \textbf{c} and \textbf{e}, see Suppl. Sect.~\ref{suppl:sec_LOD_derivation}.}
        \label{fig:3}
\end{figure*}

\section{Finding the optimal soil thickness}

With our proposed transmission geometry, we asked ourselves how one should choose the soil sample thickness. Intuitively, we expected there to be an optimal value, given that a sample that is too thick will result in both the excitation spectrum and the XRF emission being attenuated in the soil before reaching the detector, and a sample that is too thin would just contain little amounts of Pb, generating minimal signal.

As the optimal thickness will depend on the Pb concentration (influencing the soil X-ray attenuation), we decided to focus on the stricter screening level guideline of 100 ppm (Fig.~\ref{fig:2}a) \cite{epa2024lead}. We modeled the soil according to a NIST composition with density of 1.5 g/cm$^3$ (cf. Suppl. Sect. \ref{suppl:sec_soil_composition} \& Table \ref{suppl:table_soil_composition}). We then envisioned the soil sample being prepared in a thin plastic bag with negligible X-ray absorption (the plastic was included in all further simulations) and performed another 10-hour long exposure with varying sample thicknesses (Fig.~\ref{fig:2}b) with the same geometry shown in Fig.~\ref{fig:1}d. We then analyzed the Pb L-shell lines in the X-ray spectra recorded at the simulated detector  (Fig.~\ref{fig:2}c) and extracted the XRF signal and background counts in these regions.

\begin{figure*}
    \includegraphics[width=\textwidth]{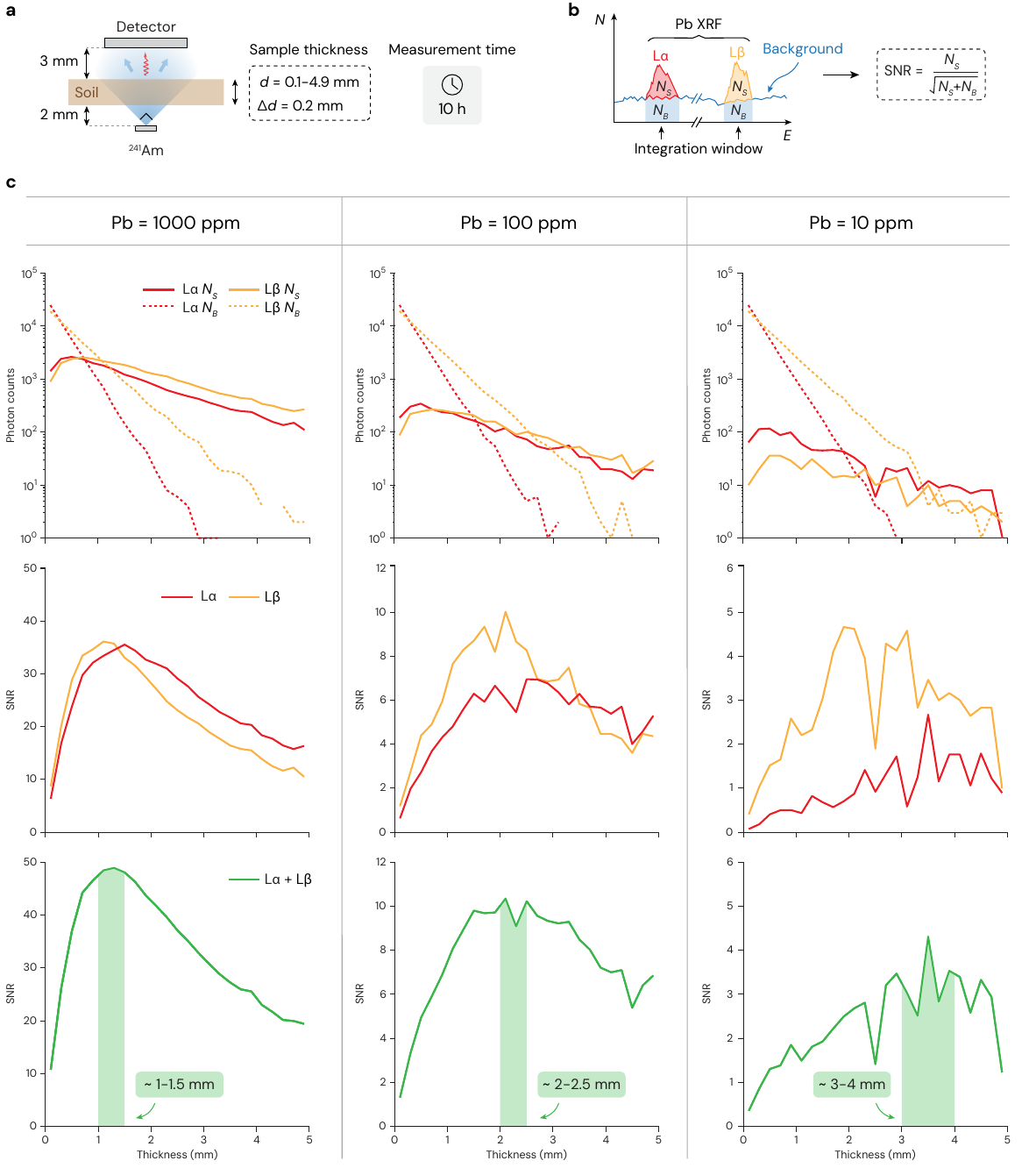}
    \caption{\textbf{Optimal sample thickness for different lead concentrations.} \textbf{a}, Transmission geometry schematic with variable thickness. \textbf{b}, Integration window definition for Pb L$\alpha$ and L$\beta$ peaks. \textbf{c}, Photon counts and signal-to-noise ratio versus thickness for 1000 ppm, 100 ppm, and 10 ppm Pb concentrations.}
    \label{suppl:fig_optimal_thickness}

\end{figure*}

We found that while the XRF signal peaks at a sub-mm soil sample thickness, with increasing thickness the sample itself filters the background faster than the XRF signal (Fig.~\ref{fig:2}d), resulting in an optimum SNR for 2--2.5 mm of soil sample thickness for our target of 100 ppm Pb  (Fig.~\ref{fig:2}e). This is a positive finding from an experimental perspective, as it is very practical to prepare a plastic bag with a few millimeters of soil. As we will show in the next section, optimizing the sample thickness for the highest SNR possible is important from the perspective of reducing the required measurement time for reliable detection. Lastly, we note that the optimal thickness is thicker for lower target concentrations, and thinner for higher target concentrations (Supplementary Fig.~\ref{suppl:fig_optimal_thickness}).

\begin{figure*}[t!]
        \includegraphics[width=\textwidth]{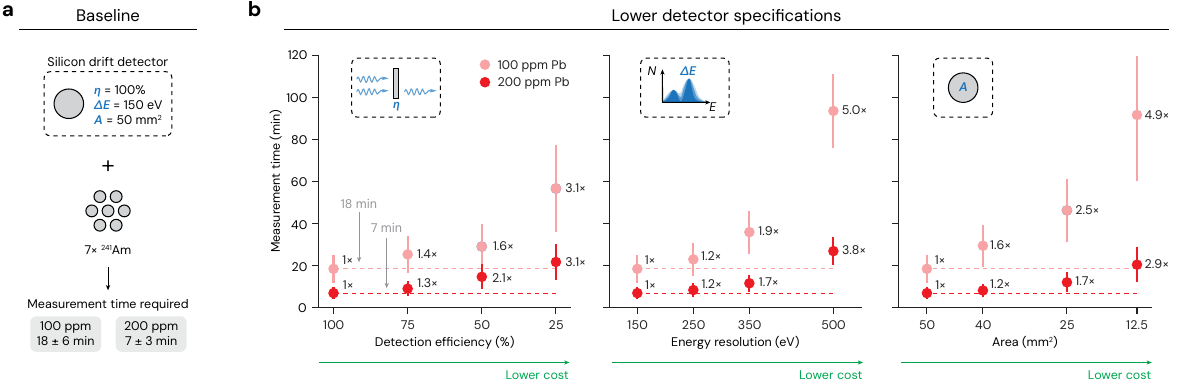}
        \caption{\textbf{Requirements on the detector.} \textbf{a}, Baseline detector configuration with 100\% efficiency ($\eta$), 150 eV resolution ($\Delta E$), and 50 mm$^2$ area ($A$), similar to silicon drift detectors commonly found in commercial handheld XRF devices. \textbf{b}, Measurement time scaling factors (shown as $\times$) for reaching a limit of detection (LOD) of 100 and 200 ppm Pb when independently varying detection efficiency (25--100\%), energy resolution (150--500 eV), and active area (12.5--50 mm$^2$).}
        \label{fig:4}
\end{figure*}


\section{Measurement time for soil screening}

Having established the optimal soil thickness, we next wanted to explore an optimized setup to achieve practical measurement times for screening applications (targeting a $\sim$100 ppm Pb limit of detection). The low fluence from a single $^{241}$Am source naturally led us to consider using multiple sources. We designed a hypothetical setup with seven $^{241}$Am sources from smoke detectors arranged around a silicon drift detector with 50 mm$^2$ active area (Fig.~\ref{fig:3}a). Since the $^{241}$Am component in smoke detectors is typically smaller than 5 mm in diameter, this proposed arrangement should be practical to prototype.

We then simulated soil samples with Pb concentrations ranging from 10 to 1000 ppm over extended exposure times. Representative spectra acquired at 10, 30, and 60 minutes (Fig.~\ref{fig:3}b) demonstrate how the Pb L$\alpha$ and L$\beta$ peaks emerge from background as measurement time increases. We quantified the signal-to-noise ratio of the Pb XRF signal by combining both L$\alpha$ and L$\beta$ peaks and plotting SNR against measurement time for all concentrations (Fig.~\ref{fig:3}c). These SNR curves follow the expected square root relationship with time, characteristic of photon counting statistics (Suppl. Sect. \ref{suppl:sec_LOD_derivation}).

To determine the the limit of detection (LOD), we decided to use the conventional 3$\sigma$ criterion \cite{currie1995nomenclature}, equivalent to an SNR>3 requirement for considering a signal reliably detectable (Fig.~\ref{fig:3}d). With this criterion, we extracted the LOD as a function of measurement time (Fig.~\ref{fig:3}e) from our SNR data (Fig.~\ref{fig:3}c). For a single $^{241}$Am source, detecting 100 ppm Pb would require 82$\pm$29 minutes, while detecting 200 ppm requires 29$\pm$11 minutes. When using the seven-source configuration (Fig.~\ref{fig:3}a), the required measurement times decrease substantially to 18$\pm$6 minutes for 100 ppm and 7$\pm$3 minutes for 200 ppm. This demonstrates that our proposed transmission XRF concept with multiple $^{241}$Am sources is a promising path towards detecting lead at relevant soil screening levels within practical measurement times (<30 min).


\section{Requirements on the detector}

Having demonstrated that our proposed transmission XRF concept is a promising approach for replacing the expensive and technologically complex X-ray tube with accessible $^{241}$Am sources, we now turn our attention to the detector component. Modern silicon drift detectors are commonly used in XRF devices for their excellent spectroscopic performance, but these are very expensive (>\$5000). Given our goal of creating a low-cost device, we investigated the requirements of the detector and how its specifications influence measurement time to identify opportunities for cost reduction. We systematically varied three key detector parameters: detection efficiency, energy resolution, and active area.

For the baseline comparison, we used the 7$\times$ $^{241}$Am configuration with a large-area SDD (Fig.~\ref{fig:4}a, same as in Fig.~\ref{fig:3}), resulting in the baseline measurement times to reach the 100 and 200 ppm Pb screening target levels shown in Fig.~\ref{fig:3}e. We then independently varied each detector parameter while holding the others at baseline values (Fig.~\ref{fig:4}b). We found that the superb detection efficiency of SDDs at the $\sim$10 keV range relevant for Pb XRF detection is not a strict requirement for cost reduction, and that a moderately degraded energy resolution can also be acceptable. However, detector area remains important as it directly relates to the fraction of isotropically emitted XRF signal detected from the soil sample. Reducing the detector area is therefore not a favorable cost-reduction strategy.

Although the trade-off between detector specifications and manufacturing cost is complex, the results in Fig.~\ref{fig:4}b establish a design space for guiding future development of our proposed device based on the transmission XRF concept (Fig. \ref{fig:1}b) according to specific budget and measurement time requirements.


\section{Discussion}

Community-level screening of heavy-element contamination in our environment requires scalable low-cost detection technologies, which currently do not exist. While laboratory-based spectrometry and current XRF devices can detect lead effectively, both are unsuitable for widespread deployment. This is particularly true in LMICs where contamination burden is highest yet resources are most limited. For widespread adoption by government agencies, NGOs, and communities in both high- and low-resource settings, instrumentation must be scalable and simple to maintain. Our proposed transmission XRF concept addresses these challenges by showing that expensive and technologically complex X-ray tubes (>\$5000) can be replaced with $^{241}$Am sources from ionizing smoke detectors ($\sim$\$10), despite the latter being over $10^8\times$ weaker. We believe this opens a route towards democratizing instrumentation for environmental lead screening.

The concept we present here is based on experimental measurements of the smoke detector $^{241}$Am spectrum as well as Monte Carlo simulations that accurately capture the physics of XRF signal generation in soil. However, some limitations are worth mentioning. Our detector modeling does not account for Compton scattering within the detector volume, though this effect is minimal at the relevant energies ($\sim$10-keV-range). Further, we only investigated a single soil composition with a fixed density of $\rho = 1.5$ g/cm$^3$, based on a well-characterized residential soil reference from NIST (Suppl. Sect.~\ref{suppl:sec_soil_composition}). Different soil compositions, particularly variations in matrix elements and density, influence the X-ray attenuation and thus the optimal thickness and measurement times required to reach desired detection limits. 

As our next step, we plan to prototype a device based on this transmission XRF concept and share the designs openly through the OpenXRF initiative (\href{http://www.openxrf.org}{www.openxrf.org}), with the goal of inspiring the next generation of low-cost, open-source instrumentation for environmental heavy-element detection. While we have focused on soil screening, given the well-established guidelines and comparatively high screening threshold values, the transmission XRF concept can be extended to samples such as spices, plastics, and other household items \cite{lopez2022assessing,sargsyan2024rapid}, with the primary difference being optimization of sample thickness based on the specific elemental composition and density of the matrix.


\section{Methods}

\subsection{$^{241}$Am spectrum measurement}

We extracted $^{241}$Am samples from two different brands of ionizing smoke detectors and measured their X-ray emission spectra using a silicon drift detector (50 mm$^2$ active area, SiriusSD, RaySpec) positioned 20 mm from the $^{241}$Am samples. Both samples exhibited spectra in agreement with each other. We reconstructed the emission spectrum by normalizing the measured spectrum with the detection solid angle and the detection efficiency of the SDD, yielding a total fluence of $9.95 \times 10^{2}$ photons/s/sr (Fig.~\ref{fig:1}c). Although we did not determine the exact $^{241}$Am activity, these are standard across ionizing smoke detectors around 37 kBq (or 1 $\mu$Ci). The fluence per solid angle and the spectral probability distribution were then used for the Monte Carlo simulations with Geant4.

\subsection{Monte Carlo simulations}

We based our results (cf. Fig.~\ref{fig:1}-\ref{fig:4}) on Monte Carlo simulations; the most accurate method to simulate X-ray interaction with matter. Below we outline some details on the simulations, supplemented by code available in the \href{https://github.com/OpenXRF}{OpenXRF GitHub Repository}.

\subsubsection{Geant4 framework and physics}

We performed all Monte Carlo simulations using Geant4 (version 10.7)~\cite{allison2006developGeant4,allison2016recentGeant4}. The core implementation consists of three main user-defined classes: \textit{construction}, \textit{generator}, and \textit{detector}, supplemented by necessary Geant4 base classes including \textit{G4VUserActionInitialization}, \textit{G4VModularPhysicsList}, and \textit{G4UserRunAction}. We used \textit{Messenger} classes to enable runtime adjustment of parameters through macro (\textit{.mac}) files, allowing parameter variation (e.g., geometry, soil sample lead concentration) without requiring recompilation.

For the physics model, we used the Livermore physics list via the \textit{G4EmLivermorePhysics} constructor, which provides accurate modeling of low-energy photon interactions including photoelectric effect, Compton scattering, and Rayleigh scattering in the $\sim$10--20~keV energy range relevant for our Pb L-shell XRF simulations.

\subsubsection{Source configuration}

We configured $^{241}$Am as a particle source using the \textit{G4GeneralParticleSource} module within the \textit{generator} class. We input the experimentally measured $^{241}$Am emission spectrum with corresponding fluence (Fig.~\ref{fig:1}c)  into the simulation. To account for the housing around the $^{241}$Am found in smoke detectors, which acts as a collimator, we set the emission cone to a half-angle of $45^\circ$, corresponding to an emission solid angle of 1.84~sr. We simulated different measurement durations by scaling the number of generated photons proportionally to the desired exposure time (e.g., 10 hours in Fig.~\ref{fig:1}~\&~\ref{fig:2}, and up to 120 min in Fig.~\ref{fig:3}).

\subsubsection{Measurement geometry}

We defined the geometries with the \textit{construction} class derived from the \textit{G4VUserDetectorConstruction} module. For all simulations, we positioned the $^{241}$Am source 2~mm from the soil sample surface and the detector 3~mm from the sample surface (cf. Fig.~\ref{fig:1}d~\&~\ref{fig:3}a). We chose these distances as short as possible to maximize XRF signal generation and collection, yet still experimentally realistic. For the multi-source configuration in Fig.~\ref{fig:3}a, we arranged seven identical $^{241}$Am sources in a hexagonal pattern (source to source distance of 5 mm) while maintaining the same vertical distances.

\subsubsection{Soil sample modeling}

We modeled the soil composition according to NIST Standard Reference Material 2587 (details provided in Suppl. Sect.~\ref{suppl:sec_soil_composition}), with all original lead content removed to create a baseline. We then added custom lead concentrations ranging from 10 to 1000~ppm by proportionally scaling all other elemental components to maintain a constant density of 1.5~g/cm$^3$. We then modeled a thin plastic bag containing the soil sample as a polyethylene bag (40~$\mu$m thickness on top and bottom surfaces, cf. Fig.~\ref{fig:2}a). Note that this plastic layer is included in all simulations, even when not seen in the schematics (e.g., Fig.~\ref{fig:1}d~\&~\ref{fig:3}a). We varied the soil sample thickness from 0.1 to 4.9~mm in 0.2~mm increments for identifying the optimal thickness (Fig.~\ref{fig:2}).

\subsubsection{Detector modeling}

We modeled the detector as a circular active area positioned perpendicular to the source-detector axis. Rather than simulating the complete detector physics, we recorded all photon interactions within the detector volume, storing the complete kinetic energy, spatial coordinates, and event number for each detected photon. This approach allowed us to apply detector characteristics during post-processing rather than during the computationally intensive Monte Carlo simulation phase. We output these data to \textit{.csv} files for analysis, with each recorded photon stored in its own row.

\subsubsection{Data processing and detector simulation}

We processed the raw simulation output (\textit{.csv} files) to emulate realistic silicon drift detector characteristics. First, we applied Gaussian convolution to the exact photon energy using a kernel with standard deviation $\sigma = \Delta E / 2.355$, where $\Delta E$ represents the full width at half maximum (FWHM) energy resolution. We then binned the convolved spectrum into 30~eV energy channels to create the final detected spectrum.

For the detector specification experiment (Fig.~\ref{fig:4}), we tuned the data processing parameters accordingly. To simulate reduced detection efficiency, we randomly discarded the corresponding fraction of photon events from the \textit{.csv} files before adding the energy resolution to ensure correct Poisson counting statistics. To vary the active detector area, we applied spatial filtering based on the recorded photon coordinates. We acknowledge that this idealized detector model does not account for effects such as incomplete charge collection, escape peaks, or complex background contributions from detector materials, which may influence real experimental performance.

\subsubsection{Signal extraction and analysis}

We extracted the Pb L-shell XRF signals by integrating photon counts within energy windows centered on the Pb L$\alpha$ and L$\beta$ peak positions. For each peak, we defined the integration window as twice the detector energy resolution (i.e., 300~eV for $\Delta E = 150$~eV) as shown in Fig.~\ref{fig:1}d. We separated the XRF signal from background by tracking each photon 
origin in the simulation output through dedicated flags that distinguished primary transmitted photons from secondary fluorescence emissions. We combined the L$\alpha$ and L$\beta$ signals to calculate the SNR and LOD values as described in Suppl. Sect.~\ref{suppl:sec_LOD_derivation}.


\section{Acknowledgements}
K.S. acknowledges funding from the Knut and Alice Wallenberg Foundation (KAW 2021.0320).


\section{Author contributions}
K.S. conceived the project after discussions with J.F. K.S. experimentally measured the $^{241}$Am spectrum in Fig.~\ref{fig:1}c. C.G. developed the Geant4 simulation environment. C.G. ran the simulations for Fig.~\ref{fig:3} \& \ref{fig:4}, and J.R. ran the simulations for Fig.~\ref{fig:1} \& \ref{fig:2}. C.G., J.R., and K.S. analyzed the data together. K.S. made the figures and drafted the manuscript. All authors revised the final manuscript.

\section{Data and code availability}
Code for interactive visualization of the data in this work can be found on the \href{https://github.com/OpenXRF/lead-screening}{OpenXRF github repository}.


\printbibliography[notkeyword={suppl}]


\clearpage
\section{Supplementary Material}
\beginsupplement

\section{Dose rate from $^{241}$Am}
\label{supp:dose_calc}

To assess the radiation safety of using $^{241}$Am sources from smoke detectors, we calculated the expected dose rate at a typical working distance of >10~cm. Using the gamma dose factor of 8.48$\times$10$^{-5}$~mSv/hr/MBq at 1~m~\cite{smith2012exposure} and a source activity of 37~kBq \cite{IAEA2024SmokeDetectors} gives the dose rate estimate:
\begin{equation}
D_{10\text{cm}} \approx 0.31~\mu\text{Sv/hr}
\end{equation}

For context, the average dose rate experienced by humans from natural background radiation is $\sim$0.2~$\mu$Sv/hr (2.4~mSv/year~\cite{unscear2008sources}), meaning that the dose rate from the X-ray emission of $^{241}$Am found in smoke detectors can be considered safe if integrated into a measurement device.


\section{Soil composition model}\label{suppl:sec_soil_composition}

We used the NIST Standard Reference Material 2587 as the foundation for our Geant4 soil model, with the original lead content removed to create a clean baseline composition. This certified reference material provides a realistic baseline composition for contaminated residential soils. Custom lead concentrations ranging from 10 to 1000 ppm were then added to this baseline by proportionally scaling all other elemental components to maintain a fixed density of $\rho = 1.5$ g/cm$^3$. Table~\ref{suppl:table_soil_composition} presents the complete elemental composition used in our simulations, including both measured NIST values and estimated compositions for major soil components not reported in the original certificate of analysis.

\begin{table*}
\footnotesize
\centering
\begin{tabular}{p{0.08\linewidth} p{0.12\linewidth} p{0.22\linewidth} p{0.1\linewidth}}
\toprule
{\bf Symbol} & {\bf Element} & {\bf Concentration (mg/kg)} & {\bf Weight (\%)}\\
\midrule
\multicolumn{4}{l}{\textbf{Major components (>1\%)}} \\
\midrule
\textcolor{red}{O} & \textcolor{red}{Oxygen} & \textcolor{red}{462,977} & \textcolor{red}{46.30} \\
Si & Silicon & 340,854 & 34.09 \\
Al & Aluminum & 60,290 & 6.03 \\
\textcolor{red}{C} & \textcolor{red}{Carbon} & \textcolor{red}{36,009} & \textcolor{red}{3.60} \\
Fe & Iron & 28,941 & 2.89  \\
K & Potassium & 16,286 & 1.63  \\
\textcolor{red}{H} & \textcolor{red}{Hydrogen} & \textcolor{red}{15,433} & \textcolor{red}{1.54} \\
Na & Sodium & 11,595 & 1.16  \\
\midrule
\multicolumn{4}{l}{\textbf{Minor components (0.1--1\%)}} \\
\midrule
Ca & Calcium & 9,537 & 0.95  \\
Mg & Magnesium & 6,883 & 0.69  \\
Ti & Titanium & 4,033 & 0.40  \\
N & Nitrogen & 2,572 & 0.26  \\
\midrule
\multicolumn{4}{l}{\textbf{Trace components (0.01--0.1\%)}} \\
\midrule
P & Phosphorus & 998 & 0.10  \\
S & Sulfur & 823 & 0.08  \\
Mn & Manganese & 670 & 0.07  \\
Ba & Barium & 584 & 0.06  \\
Zn & Zinc & 345 & 0.03  \\
Cl & Chlorine & 206 & 0.02  \\
Cu & Copper & 165 & 0.02  \\
F & Fluorine & 154 & 0.02  \\
\midrule
\multicolumn{4}{l}{\textbf{Ultra-trace components (<0.01\%)}} \\
\midrule
Sr & Strontium & 130 & 0.013  \\
Cr & Chromium & 95 & 0.010  \\
V & Vanadium & 80 & 0.008  \\
Ce & Cerium & 59 & 0.006  \\
B & Boron & 51 & 0.005  \\
Ni & Nickel & 37 & 0.004  \\
Li & Lithium & 33 & 0.003  \\
La & Lanthanum & 30 & 0.003  \\
Nd & Neodymium & 26 & 0.003  \\
Y & Yttrium & 15 & 0.002  \\
As & Arsenic & 14 & 0.001  \\
Co & Cobalt & 14 & 0.001  \\
Nb & Niobium & 14 & 0.001  \\
Ga & Gallium & 13 & 0.001  \\
Sc & Scandium & 11 & 0.001  \\
Be & Beryllium & 9 & 0.001  \\
Th & Thorium & 8 & 0.001  \\
Cd & Cadmium & 2 & 0.0002  \\
Yb & Ytterbium & 2 & 0.0002  \\
Hg & Mercury & 0.3 & 0.00003  \\
\bottomrule
\end{tabular}
\caption{%
\textbf{Baseline elemental composition of soil used for our Geant4 simulations}. These were based on \href{https://tsapps.nist.gov/srmext/certificates/2587.pdf}{NIST SRM 2587} (residential soil with original lead removed). Custom lead concentrations (10--1000 ppm) were added by scaling all other elements proportionally to maintain 100\% mass balance. Red entries indicate estimated compositions for elements not measured in the original NIST standard. Overall soil sample density was always kept at $\rho = 1.5$ g/cm$^3$ for our simulations.
}
\label{suppl:table_soil_composition}
\end{table*}


\section{Defining SNR and LOD}\label{suppl:sec_LOD_derivation}
Here we define the theoretical expression for the signal-to-noise ratio (SNR) and the limit of detection (LOD) to measure the concentration of lead in soil using our transmission XRF approach (cf. Fig.~\ref{fig:3}).

In our XRF measurements, the detected XRF signal comes from lead (to first approximation proportional to the lead concentration $c$ in ppm) and background from scattered incident radiation. We define the SNR for a measurement of duration $t$ (in seconds) as
\begin{equation}
\text{SNR}(c,t) = \frac{N_S}{\sqrt{N_S + N_B}}
\label{eq:snr_definition}
\end{equation}
where $N_S = s(c) \cdot t$ is the total XRF signal counts and $N_B = \beta \cdot t$ is the total background counts \cite{james2006statistical}. The signal count rate follows $s(c) = \alpha\cdot c$, where $\alpha$ can be regarded as the signal detection rate (counts/ppm/s), and $\beta$ is the background count rate (counts/s). Substituting these relationships into Eq.~\ref{eq:snr_definition}:
\begin{equation}
\text{SNR}(c,t) = \frac{s(c) \cdot t}{\sqrt{[s(c) + \beta] \cdot t}} = \frac{\alpha c}{\sqrt{\alpha c + \beta}} \cdot \sqrt{t}
\label{eq:snr_expanded}
\end{equation}
This shows that the SNR increases with the square root of measurement time, consistent with Poisson counting statistics. We use this relationship to fit the SNR as a function of measurement time in Fig.~\ref{fig:3}c.

For calculating the LOD in ppm as a function of measurement time, we adopt the standard $3\sigma$ criterion, which is equivalent to stating that SNR = 3 is required to claim the presence of lead in the soil samples \cite{currie1995nomenclature}. Inserting this into Eq.~\ref{eq:snr_expanded} yields:
\begin{equation}
3 = \frac{\alpha\cdot \text{LOD}}{\sqrt{\alpha\cdot \text{LOD} + \beta}} \cdot \sqrt{t}
\label{eq:lod_criterion}
\end{equation}
Solving for the measurement time required to reach the $\text{LOD}$ yields:
\begin{equation}
t = \frac{9(\alpha \cdot \text{LOD} + \beta)}{\alpha^2 \cdot \text{LOD}^2}
\label{eq:time_vs_lod}
\end{equation}
To express the LOD as an explicit function of measurement time, we rearrange Eq.~\ref{eq:time_vs_lod} to obtain the quadratic equation:
\begin{equation}
t \alpha^2 \cdot \text{LOD}^2 - 9\alpha \cdot \text{LOD} - 9\beta = 0
\label{eq:quadratic}
\end{equation}
Applying the quadratic formula, taking the positive root, and simplifying the expression yields:
\begin{equation}
\text{LOD}(t) = \frac{9 + \sqrt{81 + 36\beta \cdot t}}{2\alpha t}
\label{eq:lod_formula}
\end{equation}
For practical curve fitting (cf. Fig.~\ref{fig:3}e), we rewrite Eq.~\ref{eq:lod_formula} in a simplified two-parameter form:
\begin{equation}
\boxed{\text{LOD}(t) = \frac{A + \sqrt{A^2 + Bt}}{t}}
\label{eq:lod_fit}
\end{equation}
where $A$ and $B$ relate to the physical quantities as:
\begin{align}
A &= \frac{9}{2\alpha} \label{eq:A_param}\\
B &= \frac{9\beta}{\alpha^2} \label{eq:B_param}
\end{align}

\section{Measurement time uncertainty} \label{suppl:sec_sigma_t}

In counting statistics, the statistical uncertainty of a measured quantity $X$ is defined as the square root of its variance \cite{james2006statistical}. If $X$ is Poisson-distributed with a mean value of $N$, the variance equals the mean:
\begin{equation}
\mathrm{Var}[X] = N.
\end{equation}
As defined earlier in Sect.~\ref{suppl:sec_LOD_derivation}, the total signal and background counts recorded during a measurement time $t$ are $N_S = s \cdot t$ and $N_B = \beta \cdot t$, respectively. Using the property $\mathrm{Var}(a \cdot X) = a^2 \cdot \mathrm{Var}(X)$, the variances of the count rates can be expressed as:
\begin{equation}
\mathrm{Var}(s) = \frac{s}{t}, \qquad
\mathrm{Var}(\beta) = \frac{\beta}{t}
\end{equation}
This yields the statistical uncertainties
\begin{equation}
\sigma_s = \sqrt{\frac{s}{t}}, \qquad \sigma_\beta = \sqrt{\frac{\beta}{t}}.
\end{equation}
Since the required measurement time $t$ (Eq.~\ref{eq:time_vs_lod}) to reach a certain LOD is itself a function of the random variables $s$ and $\beta$, its uncertainty can be estimated using Gaussian error propagation:
\begin{equation}
\sigma_t^2 = 
\left(\frac{\partial t}{\partial s}\right)^2 \sigma_s^2 +
\left(\frac{\partial t}{\partial \beta}\right)^2 \sigma_\beta^2.
\end{equation}
From Eq.~\ref{eq:time_vs_lod}, at a given concentration $c$ (e.g., at the LOD), the required measurement time is:
\begin{equation}
t = \frac{9(s + \beta)}{s^2},
\end{equation}
where $s = s(c) = \alpha \cdot c$. The partial derivatives are:
\begin{equation}
\frac{\partial t}{\partial s} = \frac{9s^2 - 18s(s+\beta)}{s^4} = -\frac{9(s+2\beta)}{s^3}, \qquad
\frac{\partial t}{\partial \beta} = \frac{9}{s^2}.
\end{equation}
Substituting these derivatives and the rate uncertainties into the error propagation formula yields:
\begin{equation}
\sigma_t^2 = \frac{81(s+2\beta)^2}{s^6} \cdot \frac{s}{t} + \frac{81}{s^4} \cdot \frac{\beta}{t} = \frac{81}{t}\left(\frac{(s+2\beta)^2}{s^5} + \frac{\beta}{s^4}\right).
\end{equation}
Thus, the statistical uncertainty of the measurement time required to reach the LOD (cf. Fig.~\ref{fig:3}e) is:
\begin{equation}
\boxed{\sigma_t = 9 \sqrt{\frac{1}{t}\left( \frac{(s+2\beta)^2}{s^5} + \frac{\beta}{s^4} \right)}}.
\label{eq:sigma_t}
\end{equation}

\printbibliography[keyword={suppl}]

\end{document}